\documentclass[12pt,preprint]{aastex}

\shorttitle{NGC~4051 in the low state}
\shortauthors{Uttley et al.}

\begin{document}


\title{Catching NGC~4051 in the low state with {\it Chandra}}


\author{Philip Uttley\altaffilmark{1}, Antonella
Fruscione\altaffilmark{2}, Ian M$^{\rm c}$Hardy\altaffilmark{1} \and
Georg Lamer\altaffilmark{3}} 

\altaffiltext{1}{Department of Physics and Astronomy, University of
Southampton, Southampton SO17 1BJ, UK.  pu,imh@astro.soton.ac.uk}
\altaffiltext{2}{Harvard-Smithsonian Center for Astrophysics, 60
Garden Street, Cambridge, MA 02138.  antonell@head-cfa.harvard.edu}
\altaffiltext{3}{Astrophysikalisches Institut Potsdam, An der
Sternwarte 16, D-14482 Potsdam, Germany.  glamer@aip.de}


\begin{abstract}
We report the results from a {\it Chandra} TOO observation of the
low-luminosity Narrow Line Seyfert~1 galaxy NGC~4051, obtained during
a 6-week duration low-flux state in 2001 February.  During the {\it
Chandra} observation, the 2-10~keV source flux was
$7\times10^{-12}$~erg~cm$^{-2}$~s$^{-1}$, corresponding to a 2-10~keV
luminosity of $\sim8\times10^{40}$~erg~s$^{-1}$.
We confirm the absence of strong extended soft X-ray emission in NGC~4051, and that the low
state spectrum is dominated by the central point source.  The X-ray
spectrum has an unusual, hybrid shape, very soft below $\sim3$~keV
(and which we model with a black body of temperature 0.14~keV) and
very hard at higher energies (power-law slope of
$\Gamma\sim1$).  The lightcurves in both soft and hard bands are
significantly variable and correlated, implying a connection between hard and
soft components and proving that the hard component is dominated by primary
continuum emission and is not due to pure reflection from distant cold
material.  However, a comparison with {\it RXTE} data obtained during
the same 2001
low state suggests the presence of unusually prominent disk reflection
features, which may help to explain the apparent upwards curvature and extreme
hardness of the {\it Chandra} spectrum above a few keV. The shape of the 2001 low state
spectrum is consistent with that observed in a brief ($<3$~d) low flux
excursion during the source's normal state in April 2000, suggesting that emission
processes during the low state are not significantly different to
those in the normal state.  The unusual spectral shape observed in the
low state may be a continuation to low fluxes of the normal, flux-dependent spectral
variability of the source.
\end{abstract}




\section{Introduction}
NGC~4051 is a nearby ($z=0.0023$), low luminosity ($L_{\rm
X-ray}\sim3\times10^{41}$~erg~s$^{-1}$) AGN, optically classified as a
narrow-line Seyfert~1 (NLS1).  In common with other members of the
NLS1 class, and following the well-known anti-correlation of X-ray
variability amplitude with luminosity (e.g. Green, M$^{\rm c}$Hardy \&
Lehto 1993; Nandra et al. 1997), 
NGC~4051 typically shows large amplitude X-ray variability on
short time-scales
(e.g. Lamer et al. 2003; M$^{\rm c}$Hardy et al.,
in prep.).  In the medium X-ray band (2-10~keV), the X-ray continuum
slope is well-correlated with flux, with photon index ($\Gamma$)
increasing from $\Gamma\sim1.6$ to 2.4 over a
factor $\sim7$ increase in observed flux
\citep{lam03}.  Flux variations in the EUV and 2-10~keV
bands are also very well-correlated \citep{utt00a}, but a recent {\it
Chandra} High Energy Transmission Grating (HETG)
observation of NGC~4051 \citep{col01} suggests this result
is due to a correlation between the X-ray power-law continuum
and a `soft excess' which dominates in the soft X-ray band, rather
than variations in the X-ray power-law alone.

Besides the short-term X-ray variability, NGC~4051 also shows
long-time-scale flux changes, as revealed by monitoring with the {\it
Rossi X-ray Timing Explorer} (RXTE).  In particular, on two occasions -
in 1998 and later in 1999 - NGC~4051
appeared to enter a distinct, prolonged low-flux state which
lasted 5 and 3 months respectively \citep{utt99,utt00b}, during which
the absolute amplitude of X-ray variability
was much smaller than the large amplitude variations normally
seen.  The X-ray spectrum of NGC~4051 during
this `low state', was measured simultaneously by {\it BeppoSAX} and
{\it RXTE} in May~1998 \citep{gua98,utt99} and 
found to be extremely hard
($\Gamma\sim0.8$) above
$\sim4$~keV, also showing a prominent ($EW\sim1$~keV) iron K$\alpha$
line. Below 4~keV, the spectrum was found to be extremely soft 
(with a power-law fit giving $\Gamma\sim3$). Flux variability in
either the soft or hard band was
not detected during these $\sim$day-long observations.  The natural
interpretation of the low state was that the central X-ray source had
somehow switched off, revealing the hard spectrum and prominent
fluorescent iron K$\alpha$ line expected from distant
($>5$~light-months) cold material,
possibly the molecular torus, which did not have time to respond to
the long-term change in flux \citep{gua98,utt99}.  The soft X-rays
might also correspond to some extended component, a view supported by
a claim by \citet{sin99} that {\it ROSAT} HRI observations of NGC~4051
showed evidence for extended ($\sim10$~arcsec) soft X-ray emission, of
a comparable flux to that observed during the low state.

The hypothesis that the low states in NGC~4051 represent some physically
distinct behaviour of the AGN, possibly with the central X-ray
source switching off, rather than merely a low-flux epoch associated
with the same variability processes which produce the short-term
variability, is supported by optical monitoring observations. 
\citet{pet00} have shown that during the 1998 low state,
the variable component of the He~{\sc II} 4686\AA\ emission line
seems to disappear while H$\beta$ remains strongly variable,
implying that the variable component of the ionizing continuum above
$\sim54$~eV disappears during the low state, possibly due to the
transition of the inner accretion disk to a radiatively inefficient
advectively dominated accretion flow (ADAF).

In order to test models for the origin of the low states in NGC~4051,
we were awarded a 50~ks target of opportunity observation (TOO) using
the ACIS-S instrument on {\it Chandra}, to measure
the continuum spectral shape of the primary X-ray source
spectrum over a broad band, and disentangle the soft continuum of the
central source from any extended soft emission.  NGC~4051 entered into a low state of
$\sim2$~months duration in January 2001, and was observed by {\it
Chandra} in early February, as described below.

\section{Observation, data reduction and analysis}
\begin{figure}
\epsscale{1.0}
\plotone{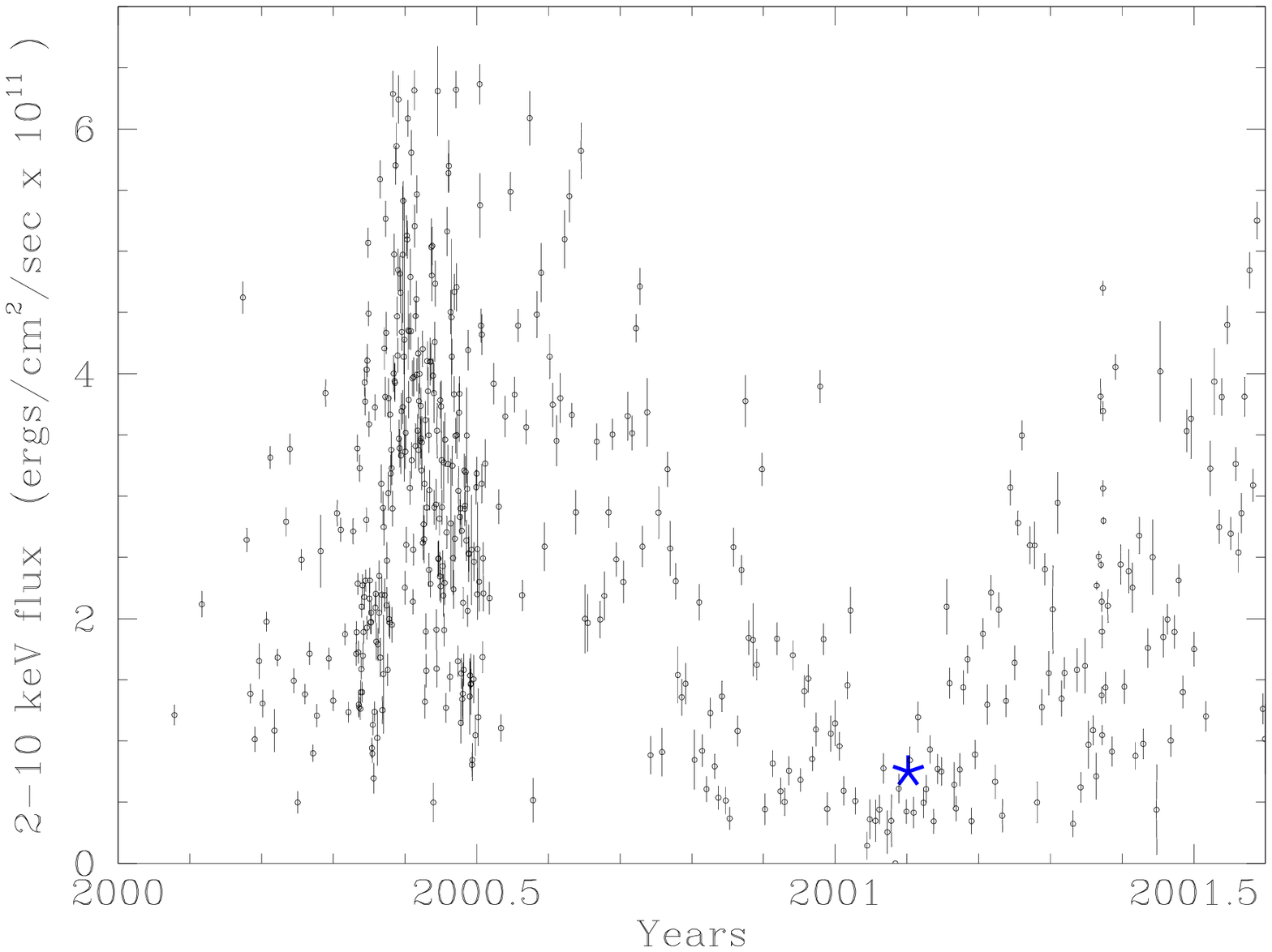}
\caption{Long-term 2-10~keV lightcurve of NGC~4051, measured by {\it
RXTE}.  The {\it Chandra}
observation (and corresponding 2-10~keV flux) is marked by a blue star.} \label{monlc}
\end{figure}

The {\it Chandra} TOO of NGC~4051 was triggered using data from our
continuing long-term monitoring observations of NGC~4051 with {\it
RXTE} (Fig.~\ref{monlc}).  In late 2000, NGC~4051 showed a decline in
long-term average X-ray flux over several months, before entering a prolonged
low-flux state around 2001~Jan~10.

\subsection{Data reduction}
{\it Chandra} successfully observed NGC~4051 on 2001~Feb~6
06:24:59-20:55:58~UT, for a total useful exposure of $\sim46$~ksec with
the spectroscopic array of the ACIS instrument (ACIS-S) without any
transmission grating in place. The source was located on the
back-illuminated chip
(S3) and offset by the standard ~20\arcsec\ from the
default on-axis location to avoid a node boundary (see the {\it Chandra} Proposers
Observatory Guide [POG]). 
We chose a 1/8 subarray mode with 0.43 s frametime to mitigate
pile-up effects,
resulting in a 1\arcmin\ x 8\arcmin\ image.   The analysis of the data
was performed using standard CIAO tools (version 2.2.1). We
regenerated the level 2 event file corrected with the latest gain and
aspect data.  We extracted the spectrum of the source from a 12 pixels
radius (~6\arcsec) circular aperture, to yield a total of 57760
counts.  Assuming the empirical spectral model which we use later
(Section~\ref{empmod}), we estimate from Web-PIMMS 
a pileup fraction of $\sim14$\%.  The background spectrum was
extracted from a ~7000 pix$^{2}$ rectangular region, away from the target
and containing no obvious sources.
Because the ACIS QE
degradation\footnote{http://cxc.harvard.edu/cal/Acis/Cal\_prods/qeDeg/index.html}
affects
our data, we used the `corrarf' code provided by the {\it Chandra} X-ray
Center to correct all the effective areas used in our spectral
analysis.  Throughout the paper, we use XSPEC~v11.0.1 to carry out
spectral fits.

\subsection{Extended emission in the {\it Chandra} image}
\label{image}
\begin{figure}
\epsscale{1.0}
\plotone{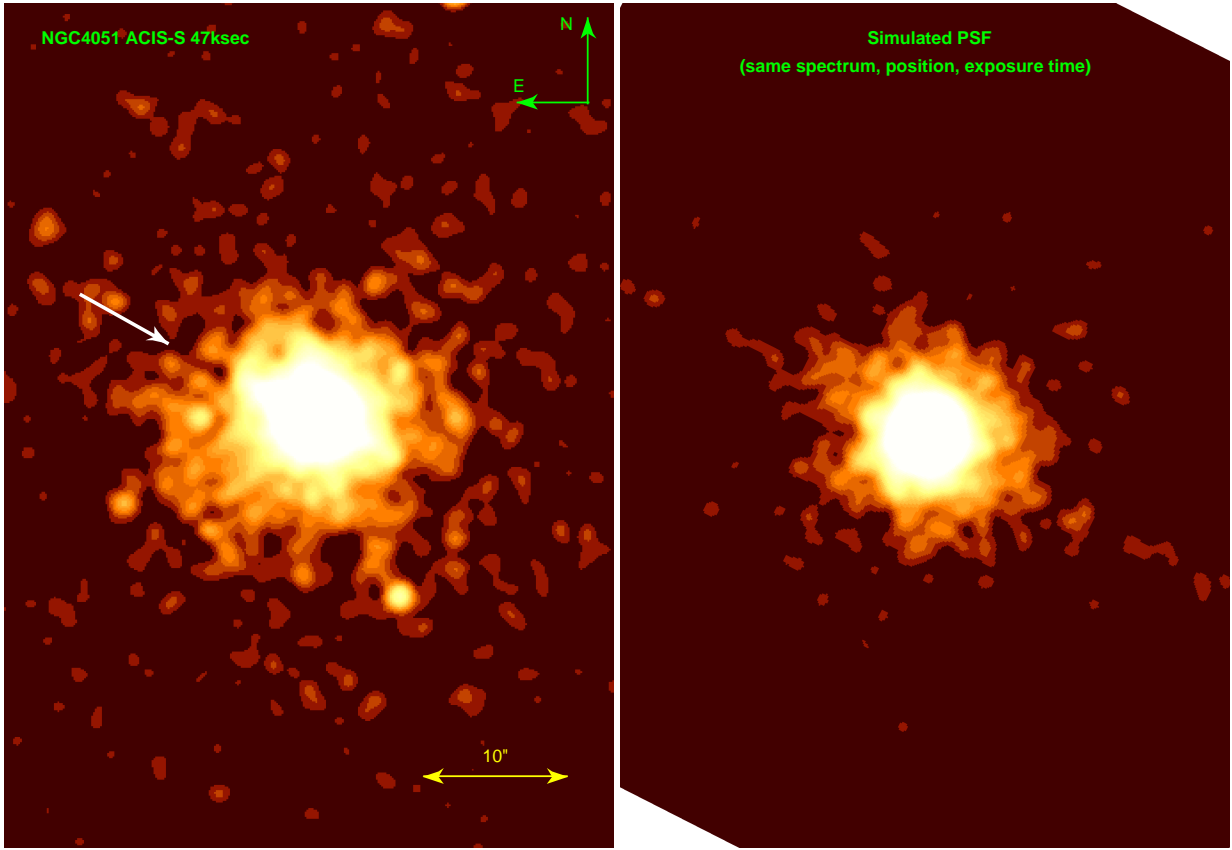}
\caption{Comparison of the Gaussian ($\sigma = 0.5\arcsec$) smoothed image of
NGC4051 and a smoothed image of a ChaRT-simulated Chandra PSF obtained
assuming the same position, spectral shape and exposure time of the
target. The two images are normalized and the logarithmic color scale is
identical. The arrow marks the direction of the removed ACIS-S
instrumental readout streak.} \label{psf} 
\end{figure}

To detect any possible weak X-ray extended emission, we extracted
and smoothed the image of the source and compared it with a simulated
Chandra PSF calculated with ChaRT\footnote{http://cxc.harvard.edu/chart} at the exact
location of the source (assuming our best
fitting empirical spectral model described later in
Section~\ref{empmod}).  The observed and simulated images are shown in Fig.~\ref{psf}. 
Fitting the azimuthally averaged
source radial profile with the simulated PSF radial profile indicates
only weak extended emission in the SE direction out to ~20\arcsec.
Details of nearby source detections and further imaging analysis will be presented in a
forthcoming paper (Fruscione et al. in preparation), so we do not
carry out any detailed analysis of extended emission here.  However, we note that
excluding the readout streak, we find 796 counts in the annular region
12-25 pixels (6\arcsec-12.5\arcsec) from the central source, which is
dominated by the central PSF so that any extended emission contributes $<1$\% of the total
observed counts, and is intrinsically very weak (i.e. much weaker than
the component claimed by Singh 1999).  This result strengthens the
finding of \citet{col01}, based on the zeroth order HETG grating
image, that there is no significant extended soft X-ray emission in
NGC~4051, and implies that the central source spectrum we extract is
dominated by nuclear emission and not contaminated by any extended
component.

\section{Broadband spectral shape and lightcurves}
\label{broadlcs}
\begin{figure}
\epsscale{1.0}
\plotone{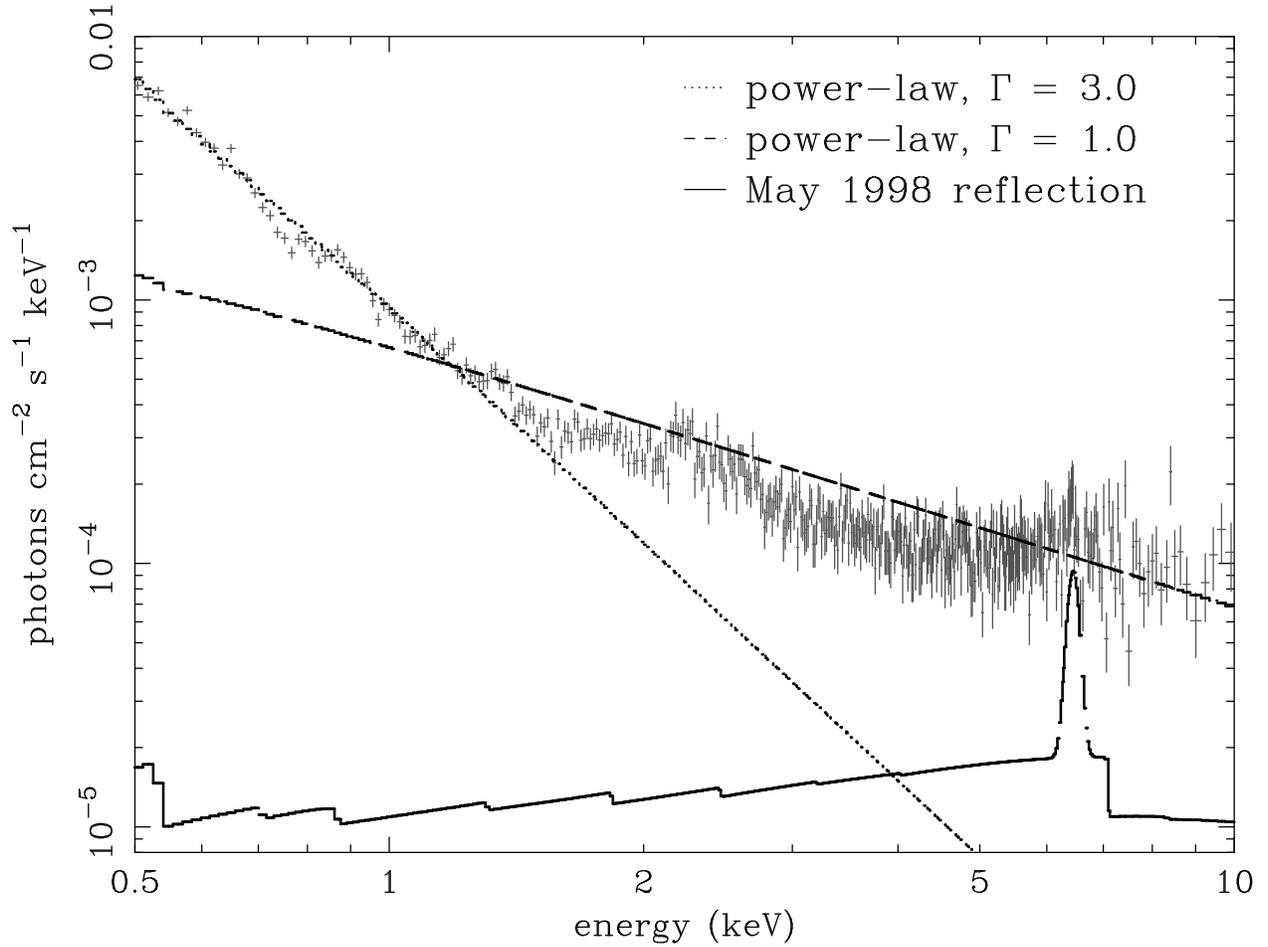}
\caption{Unfolded broadband {\it Chandra} spectrum of NGC~4051 in the
2001 February low state, compared with simple power law models and the
`reflection' model component of the 1998 May low state.} \label{ufspec} 
\end{figure}

To demonstrate the broadband shape of the {\it Chandra} 2001 low
state spectrum relative to different power-law slopes,
we first plot an {\it unfolded} spectrum (Fig.~\ref{ufspec}) fitted to a simple
absorbed power law of fixed photon index $\Gamma=1$ and Galactic absorbing column,
$N_{\rm H}=1.31\times10^{20}$ (from Elvis, Lockman \& Wilkes 1989).
The $\Gamma=1$ power-law model is plotted as a dashed line and for
comparison we plot a steep power law model ($\Gamma=3$) as a dotted line.
We also plot the model of a reflection spectrum (including iron K$\alpha$
line) that provided a good fit to the 1998~May low state spectrum
above 4~keV \citep{utt99}.   Note that although unfolded X-ray spectra
are always model-dependent, that dependence is weak for broadband
continuum models of the type we plot here, so that in the 0.5-10 keV range the
unfolded data relative to all three models is almost identical.  
Therefore the unfolded observed spectrum is a good
representation of the true continuum shape.  Clearly the 2001 low state spectrum has 
the same `hybrid' form of a very hard component above a few keV, and a
soft component at low energies, as seen in the 1998 low state.  However,
the normalization of the hard component is a factor $\sim5$
greater than in 1998~May, suggesting that the primary continuum itself
is intrinsically hard.

In Fig.~\ref{chanlcs} we show the {\it Chandra} lightcurves in soft 
(0.3-2~keV) and hard (3-10~keV) bands, together with the corresponding
hardness ratio.  Clearly the soft and hard lightcurves are well
correlated (CCF analysis shows any lag must be less than $\sim1$~ks),
but the hard band is significantly more variable than the soft band
(hard band fractional rms 29\% versus 16\% in the soft band),
so that the broadband continuum becomes harder with increasing flux.
The large-amplitude continuum variations in the hard band confirm that
we are seeing the primary continuum during the 2001 low state, and
not simply emission from a distant reflector.

\section{Spectral fits and comparison with previous observations}
\begin{figure}
\epsscale{1.0}
\plotone{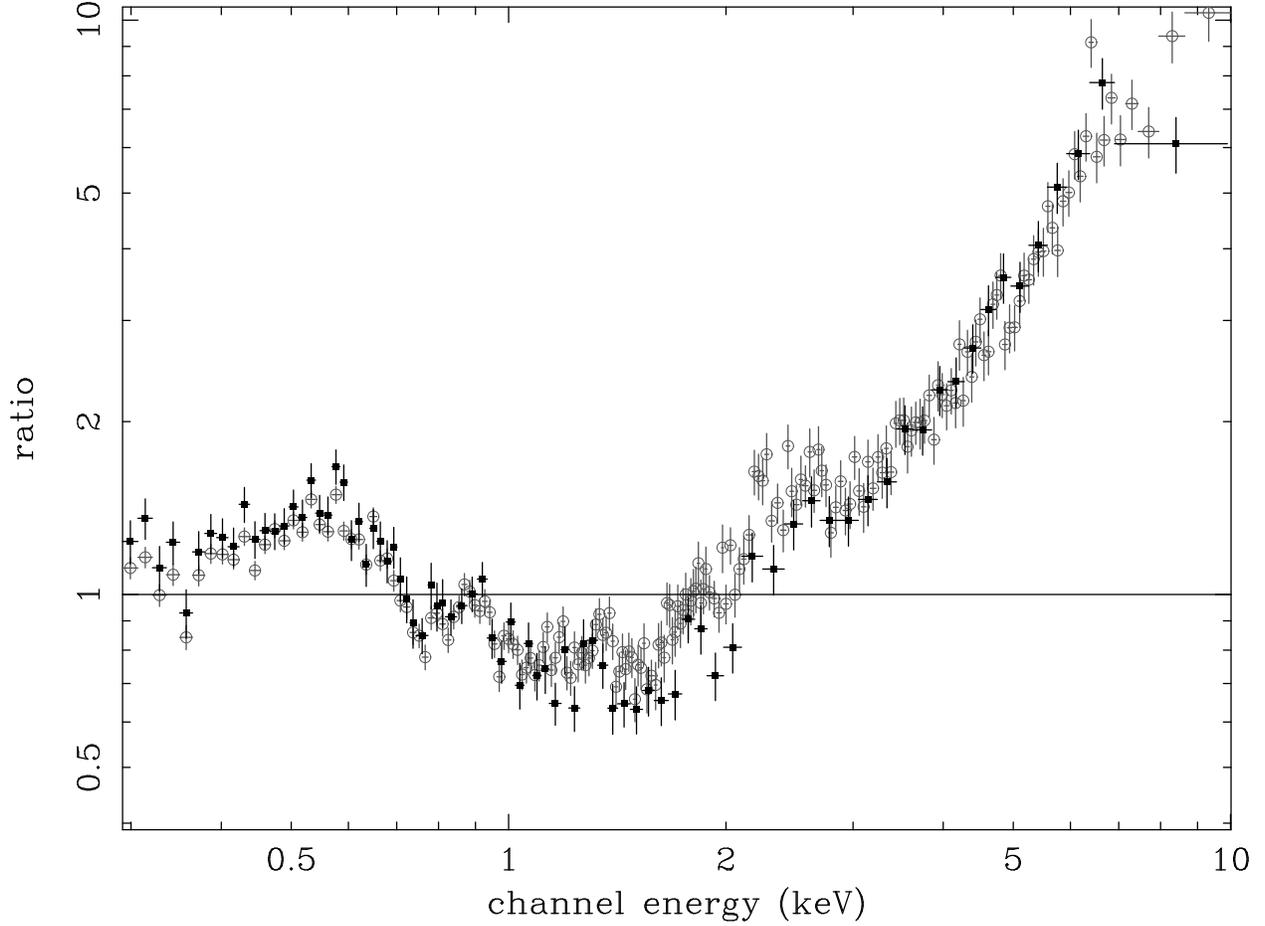}
\caption{Comparison of total spectrum (open circles) and that extracted from PSF wings
only (filled squares), with a simple power law model.  For clarity,
both spectra are binned up to have $>100$ counts per bin.} \label{plup} 
\end{figure}

The shape of the broadband spectrum of NGC~4051 in the low state is
clearly an unusual hybrid of an extremely hard component above a few keV
and a prominent soft excess at lower energies.  We can confirm that this
unusual shape is not caused by pileup effects (which would redistribute
low energy photons to higher energies and so harden the spectrum) by
comparing the spectrum with a spectrum extracted from only the wings of the
PSF (excluding counts from a circle of 1.5" diameter centred on the
source).  In Fig.~\ref{plup} we plot the ratio of the total spectrum and
PSF wing spectrum to the same simple power-law (renormalized by 0.223, the ratio
of counts in the two spectra, to fit the PSF wing spectrum).  The effect
of pileup is to harden the soft component of the spectrum somewhat, but the
discrepancy is small, consistent with the estimated pileup fraction
of 14\%, and should not affect the broad interpretation of the spectral shape,
although we should be cautious about assigning too much significance to the
specific spectral models and parameters which we fit.  With this caveat in
mind, we proceed with more detailed spectral fits to empirically
describe the spectral shape and variability.  Unless otherwise stated,
fits to the broadband {\it Chandra} spectrum are carried out to data
in the 0.5-10~keV range, in order to mitigate effects of pileup and
uncertainties in calibration below 0.5~keV. 

\subsection{An empirical model for the low state spectrum}
\label{empmod}
\begin{figure}
\epsscale{1.0}
\plotone{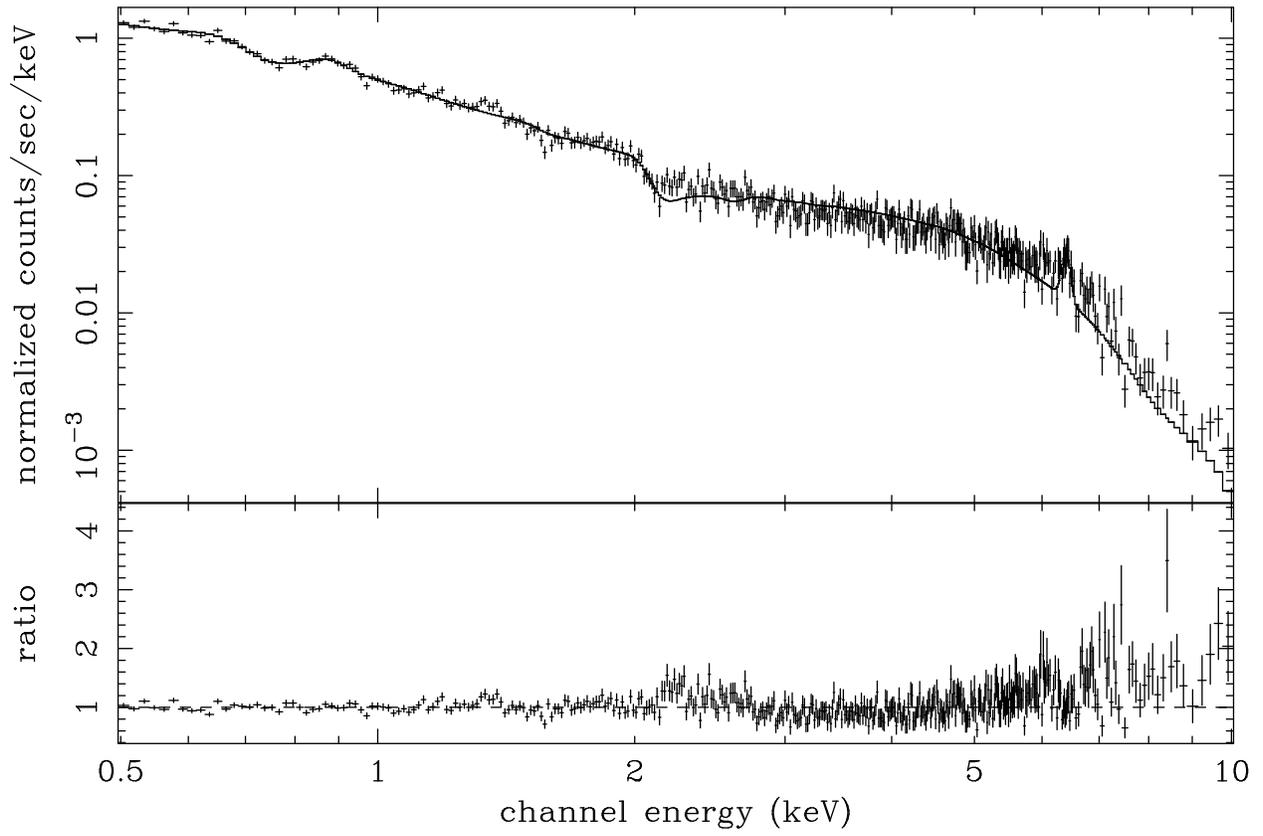}
\caption{Comparison of {\it Chandra} low state spectrum with
best-fitting model described in Section~\ref{empmod}} \label{totfit} 
\end{figure}

The 2001~February low state spectrum in the 3-10~keV band can be
well-fitted (reduced $\chi^{2}$, $\chi^{2}_{\nu}=1.06$ for 259 degrees of freedom) with
a hard power law ($\Gamma=0.54\pm0.05$) plus an
unresolved Gaussian at 6.4~keV, with equivalent width $130\pm30$~eV
(absolute line flux $(1.26\pm0.3)\times10^{-5}$~photon~cm$^{-2}$~s$^{-1}$).
Below 3~keV, the spectrum is complex (as is apparent in
Fig.~\ref{ufspec}), probably due to a combination of
emission/absorption  features, slight pileup effects on high $S/N$
data and possibly an intrinsically unusual continuum shape.  Therefore
we cannot obtain a good overall spectral fit with any model we have
attempted.  However we obtain a simple approximation to the
broadband spectral shape ($\chi^{2}_{\nu}=1.56$ for 424 degrees of
freedom) by fitting a blackbody plus power law, together
with an absorption edge at 0.7~keV (optical depth $\tau=0.63$) and an
unresolved emission line at 0.89~keV (these features likely
correspond to the 0.74~keV O{\sc vii} edge and 0.91~keV Ne{\sc ix} emission line
reported by Collinge et al. 2001).  The best-fitting flux for the
0.89~keV line feature is $2.5\times10^{-5}$~photon~cm$^{-2}$~s$^{-1}$,
consistent with that observed by Collinge et al. (2001). 
The best-fitting blackbody and power
law parameters are shown in
Table~\ref{spectab} (together with hard and soft-band fluxes) and the data, model and data/model ratio are plotted
in Fig.~\ref{totfit}.  Note that due to the statistically poor fit, we
cannot quote meaningful errors on the fit parameters.  The observed
fluxes correspond to luminosities of $\sim3\times10^{40}$~erg~s$^{-1}$
and $\sim8\times10^{40}$~erg~s$^{-1}$ in the 0.5-2~keV and 2-10~keV
bands respectively (assuming $H_{0}=70$~km~s$^{-1}$~Mpc$^{-1}$).

 The power law plus blackbody
combination leads to a steeper power law than obtained when fitting
only above 3~keV, but residuals above 5~keV
suggest that the model underestimates the hard flux, implying that
some spectral curvature may be present or some additional spectral
complexity at hard energies (Section~\ref{hardspec}).  
Replacing the blackbody with a steep
power law does not worsen the fit,
($\chi^{2}_{\nu}$ remains at 1.56 for a best-fitting power-law photon index
$\Gamma=3.4$) .  Extending the fitted spectral range to
lower energies (0.4~keV) shows an apparent flattening of the continuum
around 0.5~keV which can be replicated by the blackbody fit but not the
soft power law, although we cannot be certain that
this flattening is not due to pile-up effects.  However,
\citep{col01} also find that the soft excess observed
in the `normal' state rolls over
at low energies and is more consistent with a blackbody than a power law,
so we continue with the assumption that a blackbody
provides a good approximation to the soft excess in the low state.

All the simple models we have fitted to the broadband spectrum leave
residuals above the model at energies $>5$~keV (see
Fig.~\ref{totfit}).  It is likely then that our power-law model for
the continuum at harder energies is incorrect.  One possibility is
that significant reflection is present in the spectrum, leading to
apparent upwards curvature towards hard energies.  To test this
possibility, we included a reflection component in the model.  We used
the {\sc pexrav} model in {\sc xspec} \citep{mag95}, assuming solar
abundances and fixing inclination at
30$^{\circ}$ and power-law cutoff energy at 100~keV although note that
these latter two parameters do not affect substantially the results of
the fit.  First, we assumed that the direct
power-law continuum included in the observation 
is the primary continuum to be reflected (i.e. the reflection responds
to the change in continuum flux and $\Gamma$ in the low state).  The fit is
improved substantially ($\chi^{2}_{\nu}$=1.42 for 424 d.o.f.),
with the residuals at high energies being significantly reduced, as
expected.  The best-fitting photon index is still quite hard $\Gamma=1.36$, but the
covering factor of the reflector is $R\sim7$, and this number cannot be
substantially reduced without worsening the quality of the fit\footnote{Because
the fit is still not formally acceptable, no meaningful standard errors can be given.
However we note that a reduction of covering factor to the value of
1.0 expected for a $2\pi$ steradian sky coverage of the reflector (as
seen from the continuum source), results in $\Delta\chi^{2}=55$.}.

Such a large covering factor of reflection might be expected if the
reflector is distant (and hence is still illuminated by the normal
state continuum), for example, if the reflector corresponds to the
possible pure reflection spectrum observed in the May 1998 low state,
which could originate from cold material light months or light years
from the continuum source (\citealt{gua98},\citealt{utt99}).
Therefore, we kept the direct primary continuum slope free and
fixed the illuminating continuum
slope and normalisation at 1~keV to $\Gamma=2.3$,
$A=0.01$~photon~cm$^{-1}$~s$^{-1}$~keV$^{-1}$ respectively, 
to match those in the reflection model used
to fit the 1998 May {\it RXTE} PCA plus {\it BeppoSAX} MECS spectra
\citep{utt99}, to see if that model might explain the residuals seen at hard energies in the
2001 Feb {\it Chandra} spectrum.  With $\chi^{2}_{\nu}$=1.52 for 425
d.o.f. (for observed primary continuum $\Gamma=1.49$), the best fit
is worse than the previous fit obtained by forcing
the illuminating continuum to be identical to the direct primary
continuum.  Furthermore, the fit requires a covering factor $R=3.3$,
compared to $R=1$ which was used to fit the 1998 May spectrum.  Since
we fixed the normalisation of the illuminating continuum to that
fitted to the 1998~May spectrum, a larger covering factor might correspond to a
greater time-averaged (over months) illuminating continuum flux prior
to the 2001 low state.  However in that case we would also expect a
comparably larger 6.4 keV narrow line flux which we do not observe.
Therefore the high energy residuals from the power law cannot be
self-consistently explained by the presence of a reflection spectrum
from a distant reflector. 
Due to the difficult physical interpretation of any
reflection model, we will continue to use a simple power law in our
fits to the broadband {\it Chandra} spectrum,  as an
{\it approximate} representation of the hard spectral component.

\subsection{Broadband spectral variability within the low state}
\label{specvar}
\begin{figure}
\epsscale{1.0}
\plotone{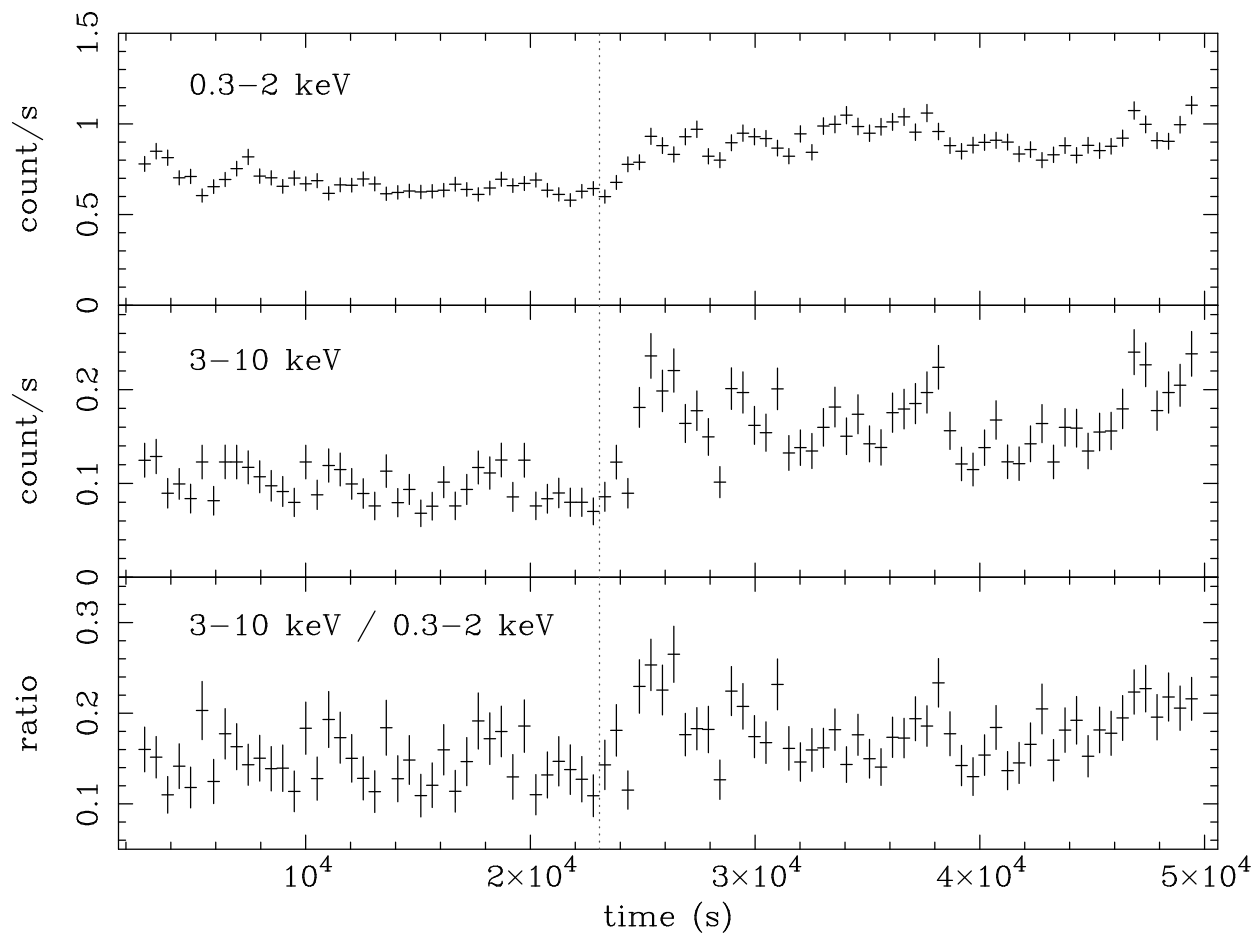}
\caption{512~s resolution {\it Chandra} 0.3-2~keV (top) and 3-10~keV
(center) lightcurves of NGC~4051 in the 2001 February low state, 
together with the corresponding hardness ratios (bottom).  The dotted
line denotes the split between high and low flux epochs used in
Section~\ref{specvar} to study
spectral variability within the low state.} \label{chanlcs} 
\end{figure}

We can split the {\it Chandra} observation into two halves of exposure
$\sim20$~ks and $\sim26$~ks respectively,
corresponding to the low and high-flux epochs seen in the lightcurves
in Fig.~\ref{chanlcs} (the split is denoted by a dotted line in the Figure),
in order to examine spectral variability in
more detail.  Fitting the simple empirical (no reflection)
model used to fit the spectrum
of the whole observation, we find that only the {\it relative} 
normalizations of the 
black body and power-law components change markedly, with blackbody and
power law normalization increasing by 28\% and 68\% respectively 
between low and high flux epochs, while the
black body temperature, power-law slope,
edge depth and line fluxes do not change noticably.  The fit is not
particularly good however ($\chi^{2}_{\nu}$=1.54 for 643 d.o.f.),
partly due to the presence of positive residuals at high energies.  The relatively small
variation in the flux of the soft spectral component compared to the hard
component is as
expected considering the lower variability seen in the soft band
(Fig.~\ref{chanlcs}).

We can also use the low and high-flux data obtained during the low
state to further test whether a constant reflection component from a
distant reflector could be present in the spectrum.  We applied the
same fixed reflection model (i.e. assuming May 1998 parameters) as
used in Section~\ref{empmod}, fixing the covering factors to be the
same in both low and high flux spectra (so the reflected component
flux is constant).  The best-fitting model again produced a better
fit than the empirical power-law model described above
($\chi^{2}_{\nu}$=1.31 for 641 d.o.f.), but requires $R\sim4$.
Furthermore, the model requires the direct power-law continuum to be
steeper at low fluxes than at high fluxes ($\Gamma=2.4$ versus
$\Gamma=1.5$), contrary to the result of Taylor et al. (2003) that the varying spectral
component hardens as flux decreases in NGC~4051, due to spectral
pivoting.  Again, errors cannot be well defined as the fit is still
not formally acceptable, but if we force the low-flux direct continuum to be flatter or
the same as the high-flux direct continuum, we obtain a best fit for
for the identical slope case (for $\Gamma=1.08$), but the 
fit is significantly worse than when the slopes are left free
($\chi^{2}_{\nu}$=1.42 for 641 d.o.f.).  The observed broadband spectral
variability is therefore inconsistent with the presence of a constant
reflection component from a distant reflector, of a magnitude which
can explain the hard residuals.  We stress however that a much weaker
reflection component (comparable to that observed in May 1998) may be
present, but it does not contribute signficantly to the hard continuum
spectral shape.

\subsection{Spectral shape above 3~keV}
\label{hardspec}
Although a simple power law plus Gaussian model provides a good fit to
the 3-10~keV {\it Chandra} spectrum (Section~\ref{empmod}), the fitted power law
is extremely hard ($\Gamma\simeq0.5$).  As noted in the preceding
Sections, combined black body plus power law
fits to the 0.5-10~keV spectrum favour a steeper (and perhaps more
plausible) photon index of $\Gamma\simeq0.9$, but show significant
residuals at energies above 5~keV (Fig.~\ref{totfit}).  As shown in
the previous Sections, these hard residuals in the broadband fits cannot be simply explained by
including a constant reflection component as expected from a distant
reflector.  It might be argued that the model is
incorrect at softer energies, e.g. if the soft component is in fact
harder than the assumed black body, then the fitted slope of the hard power law component
would steepen to compensate, resulting in an underprediction of the
flux at harder energies.  Due to the complexity of the spectral shape
at soft energies and the expected distortion of the spectrum due to
pile-up, we cannot easily determine the best model for the soft
spectral component.  Therefore, to take into account the possibility
of biases in the spectral fitting due to an incorrectly modelled soft spectrum, we now
investigate the spectral shape in the 3-10~keV range, ignoring the
data at lower energies.  

We first test the constant reflection model fitted to the broadband low and high
flux spectra in Section~\ref{specvar} (direct continuum $\Gamma$ allowed
to be free, we do not include the soft spectral components: black
body, edge and soft Gaussian in the fit).  We find an acceptable fit
($\chi^{2}_{\nu}=0.98$ for  329 d.o.f.), and hence can determine
meaningful confidence limits on the fit parameters.  We confirm the
result of the broadband fits that the hard spectral shape cannot be
largely due to reflection from a distant reflector, finding a 99\%
confidence upper limit on covering factor $R<1.7$ and corresponding
upper limits on low and high-flux direct continuum slopes of
$\Gamma=0.72$ and $\Gamma=0.6$ respectively, i.e. the varying
continuum must itself be intrinsically hard.  

An alternative possibility to explain the hard spectral shape is
that it may be associated with strong disk reflection from close to the
central black hole, which is boosted by gravitational light bending
effects (e.g. \citealt{fab03}).  In this scenario, 
the shape of at least some part of the hard residuals in the broadband
fits above 5~keV may be caused by
unmodelled emission in the form of a prominent {\it broad} iron line
and the reflection component associated with it. 
To test this possibility, we first investigated the spectrum at higher
energies using data from {\it RXTE}, which has a greater collecting
area and hence greater sensitivity to broad line features than {\it Chandra}.
We combined {\it RXTE} PCA data
from PCU2 (which is continuously switched on),
from all observations between 2001 January 10 and 2001 February 23,
using standard reduction techniques and selection criteria
(e.g. Lamer et al. 2003).  The resulting 3-15~keV spectrum
shows a prominent feature at $\sim6$~keV, which may be a strong
broad diskline.  In order to test whether such a broad line
feature might be present in the {\it Chandra} spectrum (but is not
obvious due to its broadness combined  with the low $S/N$ at high
energies) we have attempted to fit both the {\it Chandra}
3-10~keV spectra and {\it RXTE} 3-15~keV PCA spectra with the same model of a
power law, plus narrow Gaussian (adopting the best-fitting line flux
from Table~\ref{spectab}) plus Laor diskline (for a maximally rotating
black hole, Laor 1991).  We fix the diskline parameters to
be the same in both {\it Chandra} and {\it RXTE} fits, but allow the
power law normalization to vary between the fits.  We allow the power law normalization
to vary to account for the different fluxes observed during the {\it Chandra}
and combined {\it RXTE} observations.
We caution that since the source flux is particularly low, systematic errors
in the {\it RXTE} PCA spectrum may arise due to uncertainties in the
background model (e.g. see Lamer, Uttley \& M$^{\rm c}$Hardy 2000).
Therefore, the fit we obtain should be taken as merely indicative of
whether or not a broad diskline-like feature could plausibly be present in the
{\it Chandra} spectrum, rather
than an attempt to fit a realistic model (in this context, we further
note that a simple power-law is used to approximate what may in fact be a
reflection-dominated continuum).

\begin{figure}
\epsscale{1.0}
\plotone{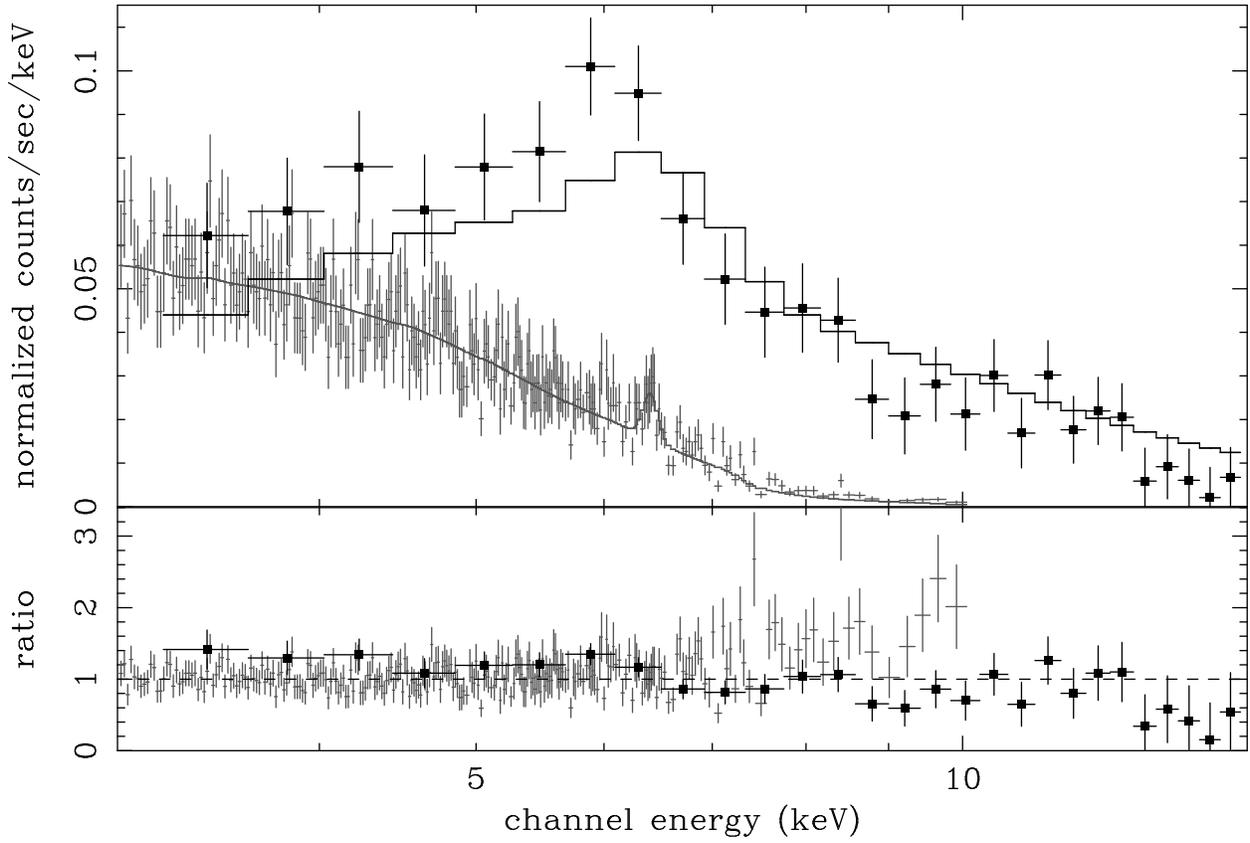}
\caption{Comparison of {\it Chandra} (plain grey crosses) and {\it RXTE}
(filled black squares) low state spectra
with a power law, plus Gaussian, plus Laor diskline model (see
Section~\ref{hardspec} for details).} \label{xtechan} 
\end{figure}

The data and best-fitting models for both spectra are plotted in
Fig.~\ref{xtechan}.  A reasonable fit is obtained to both spectra
(combined $\chi^{2}_{\nu}=1.13$ for 282 degrees of freedom) for a model with
power law slope $\Gamma=0.8$ and prominent diskline flux of
$7\times10^{-5}$~photon~cm$^{-2}$~s$^{-1}$ (corresponding to line
equivalent widths of 1~keV and 1.6~keV in the {\it Chandra} and {\it
RXTE} spectra respectively), for line energy
$E_{\rm disk}=6.9$~keV, emissivity index $\beta =-2.8$, a fixed
disk inclination angle $30^{\circ}$ and inner and outer emitting radii
of 1.23 and 33 gravitational radii respectively. 
The fit shows that the prominent broad diskline-like feature seen in the {\it RXTE}
spectrum could be present in the {\it Chandra} spectrum, and this
feature may be the cause of some of the high-energy residuals observed
in the broadband spectral fit.

Some positive high energy residuals remain in the {\it Chandra} spectrum
however, and in the opposite sense to those in the {\it RXTE}
spectrum.  It is likely that at hard energies, the {\it RXTE} PCA
spectrum may be significantly affected by uncertainties in the
background model, so we next only fit the {\it Chandra} 3-10~keV data
with a diskline including an additional reflection component to
explain the hard continuum shape in a self-consistent way.  We freeze
the diskline parameters at those values obtained in the previous fit
to the {\it RXTE} and {\it Chandra} spectra, and also include a
reflection component tied to the direct power-law continuum shape.
The strength of the reflection component is not well constrained (due
to the degeneracy with the hard power law continuum which is
required).  However, we note that a covering factor $R=3$ (for
$\Gamma=1.08$) is formally acceptable
($\chi^{2}_{\nu}=1.12$ for 260 degrees of freedom) and consistent with
the large equivalent width of the diskline ($EW\simeq 930$~eV).

\subsection{Comparison with previous observations}
\label{prevobs}
\begin{figure}
\epsscale{1.0}
\plotone{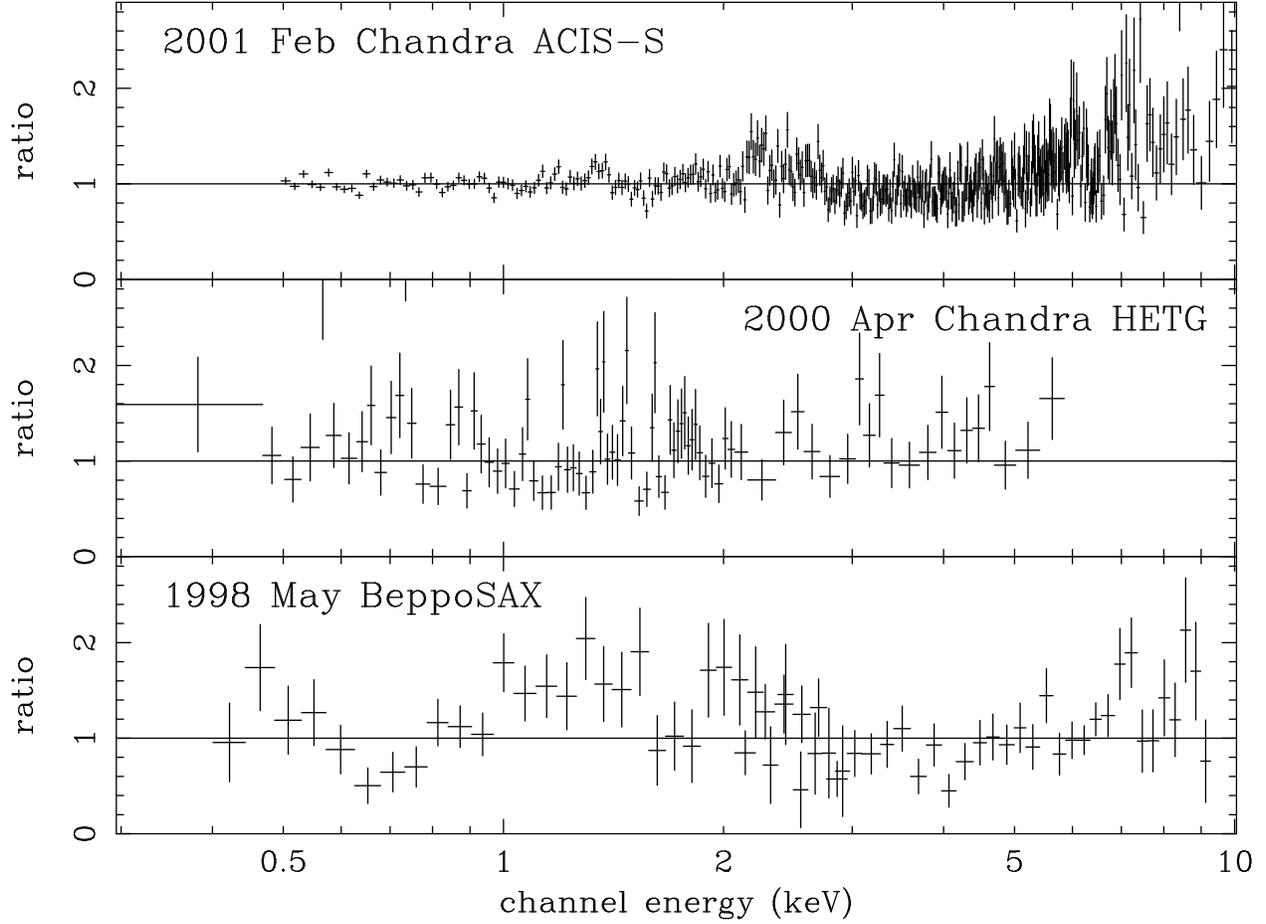}
\caption{A
comparison of the ratios of data to the best-fitting empirical model (with
free blackbody and power-law normalizations)
described in Section~\ref{empmod}
for the {\it Chandra} 2001 Feb low state spectrum (top), 
{\it Chandra} HETG 2000 Apr low flux-epoch spectrum (center)
and {\it BeppoSAX} 1998 May low state spectrum (bottom).} \label{combratio} 
\end{figure}

\begin{figure}
\epsscale{1.0}
\plotone{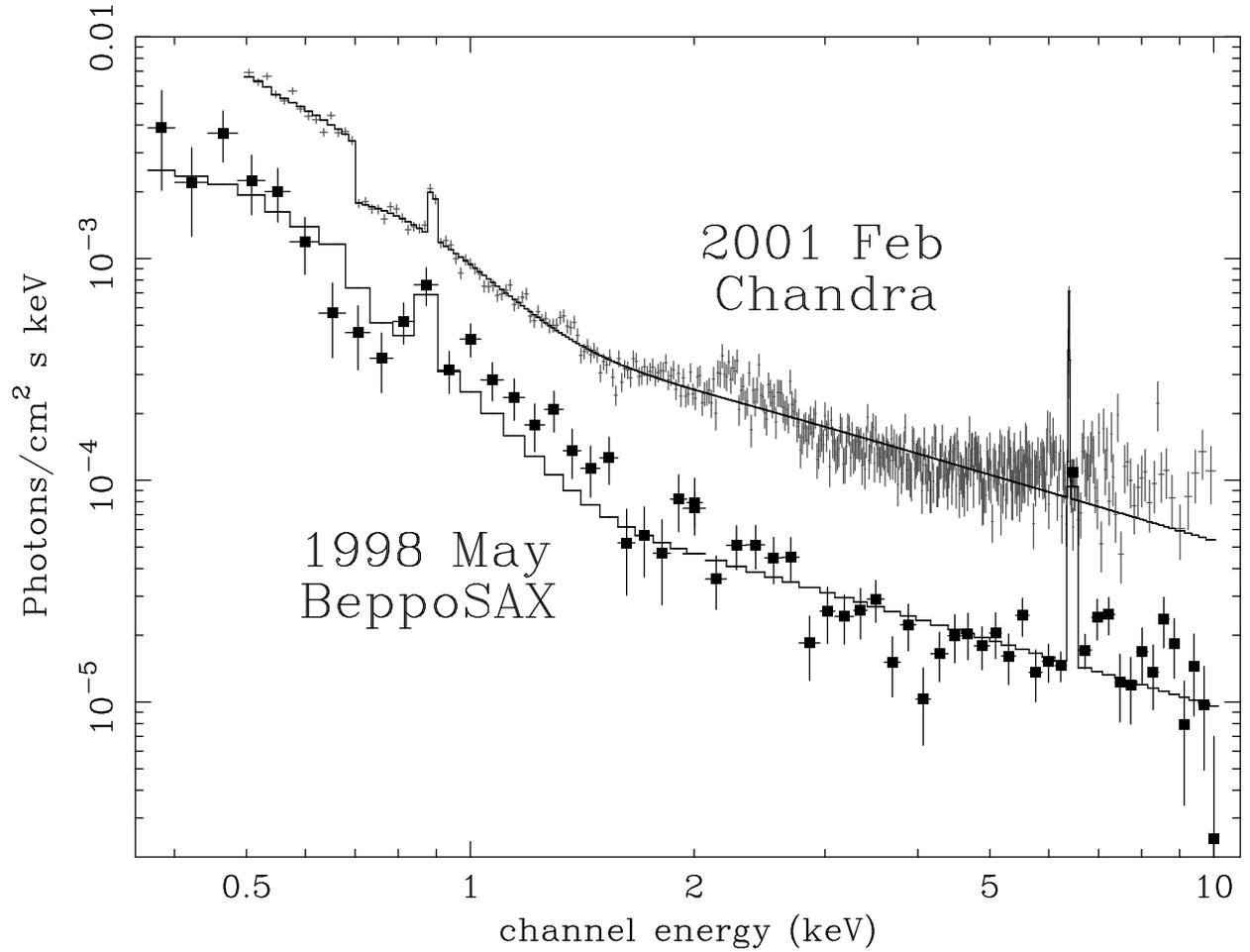}
\caption{Unfolded low state spectra for the 2001 Feb {\it Chandra}
observation (crosses) and 1998 May {\it BeppoSAX} observation, fitted to the
best-fitting model described in Section~\ref{empmod}, with free
blackbody and power-law normalizations (see also
Fig.~\ref{totfit}, Fig.~\ref{combratio}).} \label{combspec} 
\end{figure}

The broadband continuum shape of the extended low state spectrum measured by {\it
Chandra} in 2001 February is remarkably similar to that observed by
{\it Chandra}
during a brief (duration $<3$~d) low flux epoch in 2000 April,
while the source was in its normal high flux, highly variable state
\citep{col01}.  To compare the shapes of 2000 April and 2001 February spectra,
we fitted our approximate, blackbody plus power law
model to the April 2000 low flux epoch data\footnote{Corresponding to the 1st order MEG
spectrum obtained during the final 15~ksec of the April~2000
observation, grouped to $>15$~counts per spectral bin} used by
\citet{col01}.  We allowed only the blackbody and power-law 
normalizations and the depth of the 0.7~keV edge to vary.  The
data/model ratios are shown in Fig.~\ref{combratio},
together with the ratios for the fitted {\it Chandra} spectrum (Section~\ref{empmod}) 
for comparison,
and the fit parameters are shown in Table~\ref{spectab}.  We also 
carry out the same type of fit on the
{\it BeppoSAX} LECS and MECS spectra obtained during the 1998 low
state (see Fig.~\ref{combratio} and Table~\ref{spectab}).  The fit is
not very good, although the broad continuum shapes of both the 1998
low state and 2001 low states seem to match up rather well (see
Fig.~\ref{combspec}).  It is perhaps not surprising that significant
deviations from the model remain in the {\it BeppoSAX} data, since
when the source continuum flux is exceptionally low (as in 1998 May),
weak constant-flux components in the spectrum (e.g. unresolved
emission lines) are likely to be revealed.   Therefore we conclude
that, although our
simple model does not adequately fit the detailed structures in the
spectra, the same hard and soft spectral components can
approximate quite well the broadband continuum shape of both the 1998 and 2001 low
states, as well as the brief low flux epoch observed in April 2000 {\it while
the source was in a normal state}.

\section{Discussion}
\label{disc}
The X-ray spectrum of NGC~4051 during the 2001 low state is unusual, 
dominated by a very hard component ($\Gamma \sim1$) above a few keV, and
a very soft component at lower energies
($\Gamma \sim3$, although a $kT\sim0.14$~keV blackbody may be a more
appropriate model).  The question naturally arises as to what are the origins of these spectral
components.  In the 1998 low state spectrum, the hard continuum (and
prominent iron~K$\alpha$ line) above
$\sim4$~keV and the much softer continuum at lower energies (both of which did
not vary significantly), were
interpreted, respectively,  as being pure reflection from distant material and extended soft
emission \citep{gua98,utt99}.  We have seen in Section~\ref{image} that the soft emission
component in the 2001 low state is not extended in the {\it Chandra}
image, but originates from the unresolved central source.
This fact, together with a detailed spectral analysis of the {\it
Chandra} observation, show that the unusual spectrum observed in
2001 cannot be explained by the distant-reflection/extended-emission model used to explain the
1998 low state spectrum.  Instead, the emitting region must be small,
so that a substantial part of the emission can be significantly
variable on time-scales of hours, suggesting that both hard
and soft components represent mainly the primary continuum emission.
This result does not necessarily invalidate the original interpretation
of the 1998 low state spectrum, since
this earlier spectrum was obtained when the continuum was significantly
fainter than observed in February 2001, so that reflection from
distant material might have been important.  Furthermore, a
reflection model provided a significantly better fit to the 1998 low state spectrum
measured by {\it BeppoSAX} than a power-law model \citep{gua98}.
Indeed, the continuing
presence of a relatively weak, narrow Fe~K$\alpha$ line in the 2001 low state spectrum
might imply that the reflection component seen in May 1998 is present
but largely masked by the intrinsically hard primary continuum in the 2001 low state.

Having established that the 2001 low state spectrum is probably
dominated by the primary continuum emission, we must consider how this
unusual spectral shape relates to the spectrum observed during the source's
normal state.  In fact it appears (see Section~\ref{prevobs})
that the 2001 low state spectrum is not that unusual compared to the
spectrum observed at comparable flux levels during the normal state.
In other words, there may be a simple monotonic relation between
continuum spectral shape and X-ray flux which is independent of
whether the source is in a prolonged (i.e. weeks) low flux state or a
much briefer (hours to days) excursion to low fluxes.  
It is likely that the extreme hardness of the continuum at very low
fluxes is a result of the continuation of the correlation between 
spectral slope and continuum flux which is observed at higher fluxes
\citep{gua96,lam03}.   It is tempting to explain this correlation,
which is observed in other AGN (e.g. Lamer et al. 2000, Vaughan \& Edelson
2001) in
terms of thermal Comptonization models, where increases in the
seed photon flux cause corresponding increases in the hard Comptonized
flux but also cool the Comptonizing medium so that the emitted
spectrum steepens at higher fluxes and the spectrum effectively pivots
about a fixed point at high energies \citep{zdz03}.  
In fact, Taylor, Uttley \&
M$^{\rm c}$Hardy (2003) have shown using a model-independent technique
that in medium-energy X-rays (2-15~keV), the spectral variability of
NGC~4051 is consistent with a combination of spectral pivoting of a
varying power-law together with a constant hard spectral component.
It is possible that in the low state, the constant hard component is more
fully revealed, which might partly explain the deviations from a simple
power-law observed at hard energies.

We next consider how the spectrum above a few keV can be so
hard at low fluxes (irrespective of the duration of the low flux
state), and how the hard continuum relates to the soft spectral component.
The extremely hard continuum observed in the low state is difficult to 
explain using conventional thermal Comptonization models (e.g. Haardt,
Maraschi \& Ghisellini 1997), so that non-thermal models may be
required.  Alternatively, if the hard component also contains a
substantial disk-reflection component (as suggested by the {\it RXTE}
data, see Section~\ref{hardspec}), this would allow the primary
continuum to be steeper than the observed continuum, so that standard
thermal Comptonization models might still be appropriate.  It is
unclear how a disk reflection component can be so strong, unless the
disk somehow sees a much greater primary continuum luminosity than we
observe (e.g. if the continuum is beamed towards the disk through
gravitational light bending effects, Fabian \& Vaughan 2003).  However,
we note that evidence for an unusually strong iron diskline at
relatively low continuum flux levels has been reported in the Seyfert~1 galaxy
MCG-6-30-15 \citep{wil01}.  Fabian \& Vaughan (2003) have shown that
the weakly-varying hard component in the spectrum of MCG-6-30-15 may be
attributed to a strong {\it disk} reflection spectrum (see also Taylor et
al. 2003), so that at low fluxes the source should start to become
reflection dominated.  Since the normal spectral variability of NGC~4051 also implies the
presence of a constant (or weakly varying) 
hard component (Taylor et al. 2003), it is natural to infer that
strong disk reflection should be present in the low state of NGC~4051,
which may dominate over any constant reflection from the torus, even at low fluxes.

The problem of explaining the hard
continuum component in the low state of NGC~4051 is compounded by the fact that it
varies together with, and with greater amplitude than, the soft component.  If the soft
component is a source of seed photons for a Comptonizing corona (perhaps
from the hot, inner accretion disk), one
would not expect the hard, multiply scattered X-rays to vary more than the
soft photons.  The problem may be resolved if a substantial, constant
component, which does not contribute seed photons is present in the 
soft X-ray band.  The factor $\sim2$
reduction in fractional variability from the hard to the soft band can
be accounted for if half the observed soft flux arises from a constant
component (perhaps extended emission on scales $<$pc so that it is
unresolved by {\it Chandra}).  Note that the absence of varying He{\sc
ii} emission but presence of variable H$\beta$ during the low state
\citep{pet00} might be explained if, for whatever reason, the
variability amplitude in the low state is smallest in the EUV band,
but increases again towards the UV.

Although we can explain the spectral variability of NGC~4051 
within the low state by invoking a constant soft component in addition to
a variable soft component, we cannot rule out the possibility that the
spectral variability is intrinsic to the primary continuum itself. 
Furthermore, we cannot be certain that the blackbody and power-law components that
we use to model the low state spectrum really represent physically distinct
components (such as an accretion disk and corona).  It is
possible that the reason that both hard and soft components vary
together is that they have the same origin, perhaps in something like
the hybrid thermal/non-thermal plasma which can successfully fit both the
soft excess and power-law components in the spectrum of the black hole
X-ray binary Cyg~X-1 in its soft state \citep{gie99}.  

Finally we note that, since there
is no apparent difference between the low state
spectral shape and that observed during a low flux epoch in the
normal state, then the low state may not be as
special as originally thought.  This result ties in well with
the discovery (Uttley et al., in prep.) 
that the appearance of low states in NGC~4051 could be
related to the rms-flux relation which is intrinsic to the normal
variability process in AGN and X-ray binaries
\citep{utt01}.  Because of the rms-flux relation, short term variability
amplitude is correlated with long-term flux, so that
a large decrease in flux on long time-scales is also accompanied by
a corresponding large reduction in short time-scale variability
amplitude, producing the distinctive `low states'.  We therefore
consider it likely that the low states in NGC~4051 are simply prolonged low-flux
epochs in the normal life of the X-ray source, i.e. they are caused by
the same physical process which produces the flux variability on other
time-scales and do not correspond to any
distinct transitions in the physical properties of the source.  
 Although the low state of
NGC~4051 does not appear to be physically distinct from the normal
state of the source, it does allow an excellent opportunity for further TOOs to
target one extreme of the flux-correlated spectral variability in
NGC~4051, enabling further progress in understanding the
physical origin of the X-ray spectrum and variability in AGN.

\section{Conclusions}
We have presented results from a {\it Chandra} TOO observation of
the Narrow Line Seyfert~1 NGC~4051, caught in a 6-week duration low state in
2001 February.  We summarize our main results as follows:
\begin{enumerate}
\item We confirm the report of \citet{col01} that there is no strong
extended soft X-ray emission component in NGC~4051.  The soft X-ray
emission during the low state is dominated by the central point source.
\item The nuclear X-ray spectrum of NGC~4051 in the low state has an
unusual hybrid shape, very soft below $\sim3$~keV ($\Gamma\sim3$ for a simple,
power-law fit) and very hard ($\Gamma\sim1$) at
higher energies.  Both soft and hard X-ray light curves are variable and
well-correlated, the hard X-rays being more variable than the soft.
The hard X-ray variability implies that the hard continuum in the 2001
low state is not due to pure reflection from distant cold gas, but
must be the primary continuum itself.
\item We parameterize the spectrum with a simple black body plus
power-law model.  The black body model for the soft emission is
favoured over a steep power-law if one takes into account the spectral
flattening which we observe below 0.5~keV (but do not fit due to
systematic uncertainties at those energies), and is also reported by
\citet{col01} in the normal flux state.
\item The continuum at hard energies shows residuals above the
power-law model, suggesting the presence of reflection.  However the
strength of reflection required, together with an
analysis of spectral variability, shows that the deviations from a
power-law cannot be simply explained by reflection from distant
material (e.g. a molecular torus).  An alternate possibility is that the hard residuals are
caused by the presence of a strong broad feature, possibly an iron diskline and
associated reflection at higher energies.  This latter possibility is
favored when we consider {\it RXTE} data obtained during the 2001 low
state.
\item The 2001 low state spectrum is consistent with being
the same shape as the spectrum observed in a brief ($<3$~d) low-flux epoch
during the normal active state of the source, in 2000 April
\citep{col01}.  This result suggests that the unusual low state
spectrum is not a result of a fundamental physical difference in the
emission process between the low and normal states, rather it is the
continuation to low fluxes of the normal, flux-dependent spectral
variability of the source.  
\item  The even fainter 1998 May low state
spectrum is also consistent with the same hybrid model as the 2001
February spectrum. However, as the
hard component to the spectrum was a factor $\sim5$ fainter in 1998 May
than in 2001 February, and no significant variability was detected in
that earlier observation, the reflection interpretation of the 1998
May spectrum remains valid.
\end{enumerate}


\acknowledgments
We would like to thank Jon Miller and the anonymous referee
for helpful comments.  P.~U. would
like to thank the staff of the {\it Chandra} X-ray Center and the CfA for
their hospitality and support while preparing this paper.  This work has
been supported by the NASA grant GO1-2111X. 




\clearpage

\clearpage

\clearpage

\clearpage

\clearpage

\clearpage

\clearpage

\clearpage

\clearpage

\clearpage

\begin{deluxetable}{lcccccccc}
\tablecaption{Blackbody plus power law model fit parameters\label{spectab}}
\tablehead{
\colhead{Data} & \colhead{$\tau_{\rm edge}$} &
\colhead{$kT$} & \colhead{$A_{\rm BB}/10^{-5}$}
& \colhead{$\Gamma$} & \colhead{$A_{\rm PL}/10^{-5}$} & \colhead{$F_{\rm
0.5-2}$} & \colhead{$F_{\rm 2-10}$} & \colhead{$\chi^{2}/d.o.f.$} 
}
\startdata
{\it Chandra} 2001 Feb & 0.63 & 0.14 & 4.2 & 0.99 & 52.4 & 2.5 & 7.0 &
662/424 \\
{\it Chandra} 2000 Apr & 0.77 & 0.14\tablenotemark{a} & 3.6 &
0.99\tablenotemark{a} & 43.3 & 2.2 & 6.0 & 132/86 \\
{\it BeppoSAX} 1998 May & 0.52 & 0.14\tablenotemark{a} & 1.3 &
0.99\tablenotemark{a} & 9.1 & 0.7 & 1.4 & 108/60\\
\enddata
\tablecomments{Columns give $\tau_{\rm edge}$, the optical depth of
the 0.7~keV edge included in the fits; $kT$ the blackbody temperature in keV;
$A_{\rm BB}$ the {\sc xspec} blackbody normalization; $\Gamma$ the power
law photon index; $A_{\rm PL}$ the {\sc xspec} power law normalization; 
$F_{\rm 0.5-2}$ the 0.5-2~keV flux (units of
$10^{-12}$~erg~cm$^{-2}$~s$^{-1}$); $F_{\rm 2-10}$ the 2-10~keV flux
(units of $10^{-12}$~erg~cm$^{-2}$~s$^{-1}$); the $\chi^{2}$ and number
of degrees of freedom (d.o.f.) of the fit to that observation.  Note
that since none of the fits are formally acceptable, errors are not quoted.}
\tablenotetext{a}{Fixed to {\it Chandra} 2001 Feb value}
\end{deluxetable}

\end{document}